\documentclass[journal,twoside,web]{ieeecolor}
\usepackage{generic}
\usepackage{cite}
\usepackage{amsmath,amssymb,amsfonts}
\usepackage{algorithmic}
\usepackage{graphicx}
\usepackage{textcomp}
\usepackage{times}
\usepackage{epsfig}
\usepackage{graphicx}
\usepackage{amsmath}
\usepackage{amssymb}
\usepackage{adjustbox}
\usepackage{multirow}
\usepackage{arydshln}
\usepackage{comment}
\usepackage{placeins}
\usepackage[pagebackref=true,breaklinks=true,colorlinks,bookmarks=false]{hyperref}
\def\BibTeX{{\rm B\kern-.05em{\sc i\kern-.025em b}\kern-.08em
    T\kern-.1667em\lower.7ex\hbox{E}\kern-.125emX}}
\begin{document}
\title{LUNet: Deep Learning for the Segmentation of Arterioles and Venules in High Resolution Fundus Images}
\author{Jonathan Fhima, Jan Van Eijgen, Hana Kulenovic, Valérie Debeuf, Marie Vangilbergen, Marie-Isaline Billen, Heloïse Brackenier, Moti Freiman, Ingeborg Stalmans and Joachim A. Behar, \textit{Senior Member, IEEE}
\thanks{The research was supported by a cloud computing grant from the Israel Council of Higher Education, administered by the Israel Data Science Initiative. The study was performed as part of the ENRICH consortium (ERA CVD JTC 2020 project). JF, JB, and MF were supported by Grant No ERANET - 2031470 from the Ministry of Health. This research was partially supported by Israel PBC- VATAT and by the Technion Center for Machine Learning and Intelligent Systems (MLIS).}
\thanks{J. Fhima is with the Faculty of Biomedical Engineering and Department of Applied Mathematics, Technion-IIT, Haifa, Israel.}
\thanks{J. Van Eijgen, H. Kulenovic, V. Debeuf, M. Vangilbergen, M.I. Billen,  H. Brackenier and I. Stalmans are with the Research Group Ophthalmology, Department of Neurosciences, KU Leuven and the Department of Ophthalmology, University Hospitals UZ Leuven, Belgium.}
\thanks{J. A. Behar and M. Freiman are with the Faculty of Biomedical Engineering, Technion-IIT, Haifa, Israel (e-mail: jbehar.technion.ac.il).}}

\maketitle

\begin{abstract}
The retina is the only part of the human body in which blood vessels can be accessed non-invasively using imaging techniques such as digital fundus images (DFI). The spatial distribution of the retinal microvasculature may change with cardiovascular diseases and thus the eyes may be regarded as a window to our hearts. Computerized segmentation of the retinal arterioles and venules (A/V) is essential for automated microvasculature analysis. Using active learning, we created a new DFI dataset containing 240 crowd-sourced manual A/V segmentations performed by fifteen medical students and reviewed by an ophthalmologist, and developed LUNet, a novel deep learning architecture for high resolution A/V segmentation. LUNet architecture includes a double dilated convolutional block that aims to enhance the receptive field of the model and reduce its parameter count. Furthermore, LUNet has a long tail that operates at high resolution to refine the segmentation. The custom loss function emphasizes the continuity of the blood vessels. LUNet is shown to significantly outperform two state-of-the-art segmentation algorithms on the local test set as well as on four external test sets simulating distribution shifts across ethnicity, comorbidities, and annotators. We make the newly created dataset open access (upon publication). 
\end{abstract}


\section{Introduction}

The eye provides a direct view of the eye microvasculature and thereby constitutes a unique non-invasive window to the cardiovascular system. Several studies have shown that abnormalities in the retinal microvasculature can reflect on the cardiovascular health of a patient \cite{Gunn1892OphthalmoscopicTension,Scheie1953EvaluationSclerosis,Keith1939SomePrognosis,Sharrett1999RetinalStudy,Witt2006AbnormalitiesStroke}. Digital fundus imaging uses a specialized low-power microscope with a fundus camera to capture high resolution red green blue digital fundus images (DFI) of the eye’s interior surface. 
Analysis of the retinal microvasculature using DFI may allow the study of cardiovascular diseases. However, manual segmentation by an experienced annotator, of the vascular tree in a single DFI typically requires 1-2 hours. This limits large quantitative analysis of retinal vasculature changes in a certain disease, discovery of disease-specific digital vasculature biomarkers and translation of such findings to clinical practice. Modern advances in computer vision have increased the precision of segmentation in various domains. Several studies focused on DFI-based blood vessel segmentation achieved accurate results with a dice score in the range 80-89 \cite{Staal2004Ridge-basedRetina,Hoover2000LocatingResponse,Tolias1998AClustering,Jiang2003AdaptiveImages,Walter2001SegmentationTechniques,Liskowski2016SegmentingNetworks,Dasgupta2017ASegmentation,Ronneberger2015U-net:Segmentation,Guo2021Sa-unet:Segmentation,Kamran2021RV-GAN:Network}. 
However, the size of the test set in these studies was generally small with typically 10-20 DFIs and the authors did not distinguish between arterioles and venules (A/V). The case of retinal A/V segmentation remains a tedious task due the modest size of public DFI datasets with reference A/V segmentation. For example, the Digital Retinal Images for Vessel Extraction (DRIVE) dataset is the largest one and contains 40 A/V segmented DFIs \cite{Staal2004Ridge-basedRetina}. 


\begin{figure}[!tb]
    \centering
	\includegraphics[width=0.35\textwidth]{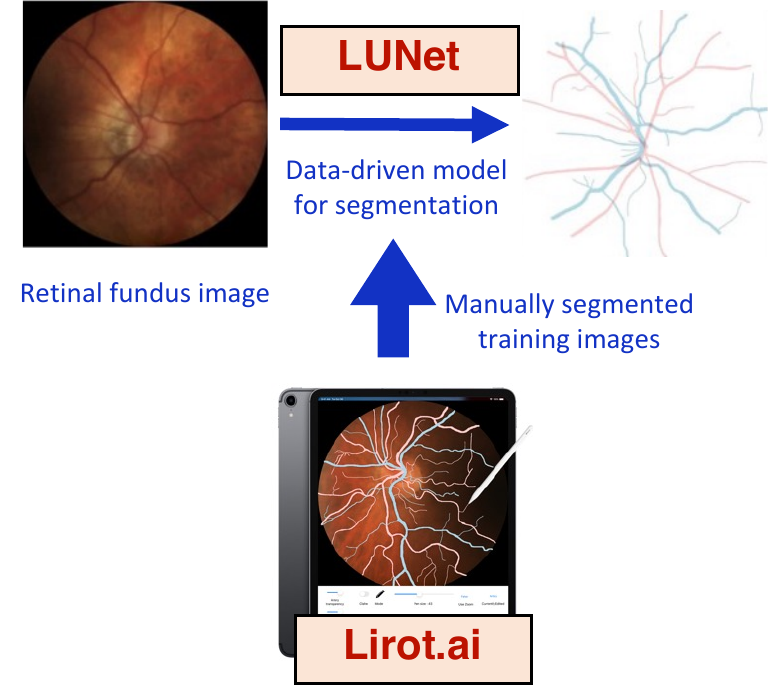}
	\caption{Overview of the experiments. DFIs were manually segmented using Lirot.ai \cite{Fhima2022Lirot.Segmentations} and used to train LUNet, a novel DL model for automatic arteriole/venule segmentation.}
	\label{fig:summary}
\end{figure}

\subsection{Prior works}
Several attempts have been made to perform automatic A/V segmentation based on DFIs. Grisan and Ruggeri used proprietary data (2003) \cite{Grisan2003AVeins} to extract vessels and classify A/V based on differences in the luminosity and contrast in the DFIs coupled with a vessel-tracking algorithm. The first use of convolutional neural networks for the A/V classification task was made by Welikala et al. in 2017, who used 100 proprietary annotated DFIs from the UK Biobank dataset as the training dataset, and the DRIVE \cite{Staal2004Ridge-basedRetina} dataset for testing \cite{Welikala2017AutomatedCohort}. 
In 2018, Hemelings et al. reported on the first use of a fully convolutional network with an auto-encoder architecture, similar to a U-Net without skip connections, to perform the A/V segmentation task \cite{Hemelings2019ArteryveinNetwork}. Two models were trained independently using the DRIVE \cite{Staal2004Ridge-basedRetina} and the High-Resolution Fundus (HRF) \cite{Budai2013RobustImages} datasets. In 2021, Hu et al. developed VC-Net, another U-Net variant incorporating a vessel constraint module to improve their model segmentation \cite{Hu2021AutomaticImages}. They trained their model on the DRIVE \cite{Staal2004Ridge-basedRetina}, HRF \cite{Budai2013RobustImages}, and LES-AV \cite{Orlando2018TowardsImages} datasets and used Tongren and Kailuan, two private datasets, for external validation \cite{Hu2021AutomaticImages}. In 2022, Galdran et al. \cite{Galdran2020TheModels} used a concatenation of two U-Net named Little W-Net to train two models for the A/V segmentation task; the first trained on the DRIVE dataset \cite{Staal2004Ridge-basedRetina} and the second oon the HRF dataset \cite{Budai2013RobustImages}. They used the LES-AV dataset \cite{Orlando2018TowardsImages} as external validation of their model trained on the DRIVE dataset. Finally, Zhou et al. developed BFN, which uses three U-Net models trained within an adversarial framework; one artery segmenter, one vein segmenter, and one multi-class segmenter \cite{Zhou2021LearningImaging}. They trained three independent BFN on the DRIVE \cite{Staal2004Ridge-basedRetina}, HRF \cite{Budai2013RobustImages}, and LES-AV \cite{Orlando2018TowardsImages} before releasing a final BFN \cite{Zhou2022AutoMorph:Pipeline} trained on these three datasets. They evaluated their final model on the IOSTAR \cite{Abbasi-Sureshjani2015Biologically-inspiredImages,Abbasi-Sureshjani2016AutomaticScores} dataset.

\subsection{Research gaps and objectives}
While previous work has focused mainly on DRIVE, HRF, and LES-AV, three public datasets that contain 40, 45, and 22 DFIs with manual reference segmentations, respectively, these datasets have different resolution, framing, field of view (FOV) and population sample. More specifically, the DRIVE dataset contains DFIs centered on the macula, with a FOV of 45$^{\circ}$ and a resolution of 584x565 pixels, acquired during a screening for diabetic retinopathy in the Netherlands. The HRF dataset contains DFIs centered on the macula, with a resolution of 2336x3504 pixels, acquired in Germany from patients with glaucoma or diabetic retinopathy and from healthy individuals. LES-AV contains DFIs centered on the optical disc, with a FOV of 30$^{\circ}$ and a resolution of 1444x1620. Patient age distribution is available for the LES-AV dataset only. The heterogeneity of these datasets led most researchers to use them to train an independent model for each dataset; poor generalization performance was observed for external datasets. Thus, there is a need for a robust A/V segmentation model with high performance that can generalize across external test sets with varying distribution shifts.

In this research, we focused on optic disc-centered DFIs, with a FOV of 30$^{\circ}$ and a high resolution of 1444x1444 pixels. In addition, a new DFI dataset, named UZLF, consisting of 240 crowd-sourced A/V segmentations performed by fifteen medical students and subsequently corrected by a senior annotator (JVE) was created. LUNet, a novel robust deep learning (DL) algorithm tailored to the A/V segmentation task, is introduced. The generalization performance of LUNet was evaluated using 30 newly manually segmented A/V DFIs from UNAF and INSPIRE-AVR \cite{Benitez2021DatasetRetinopathy, Niemeijer2011AutomatedPhotographs} as well as on the publicly available LES-AV dataset \cite{Orlando2018TowardsImages} and HRF dataset \cite{Budai2013RobustImages} which have reference A/V segmentations. LUNet was benchmarked against VC-Net \cite{Hu2021AutomaticImages} and BFN \cite{Zhou2021LearningImaging}, two open-source state-of-the-art (SOTA) algorithms.

\section{Methods}
DFIs provided by the University Hospitals of Leuven (UZ) were manually segmented using Lirot.ai \cite{Fhima2022Lirot.Segmentations}. These segmentations were used to train LUNet. Figure \ref{fig:summary} provides an overview of the experiments.

\subsection{Datasets}
A total of five datasets were used in our experiments. The University Hospital UZ Leuven Fundus (UZLF) dataset was used for developing the models while the four other datasets were used as external test sets to evaluate the generalization performance of LUNet and benchmark algorithms. The datasets are summarized in Table \ref{table:1}.


\subsubsection{University Hospital UZ Leuven Fundus (UZLF) dataset}
\begin{figure}[tb]
    \centering
	\includegraphics[width=0.5\textwidth]{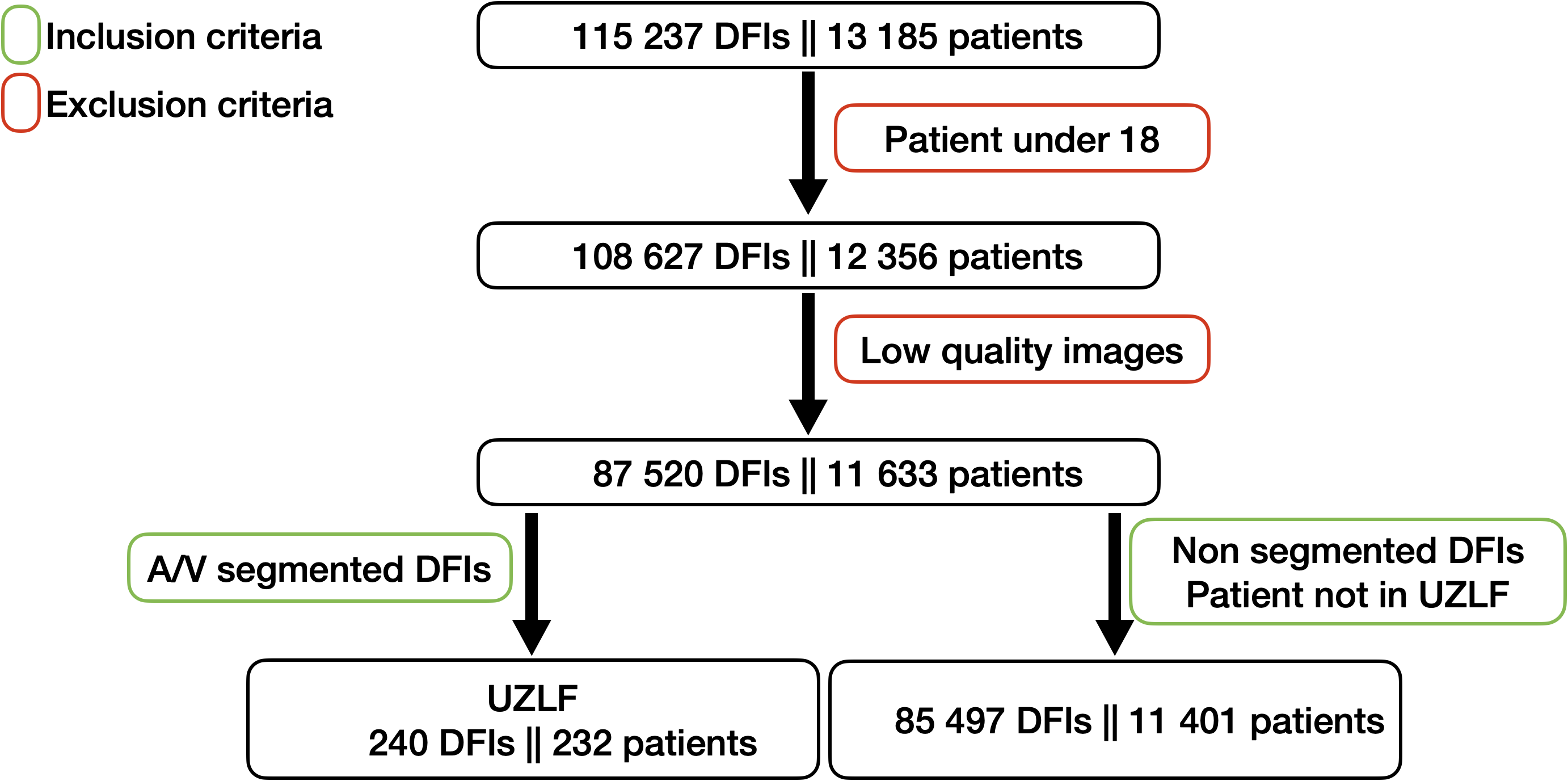}
	\caption{UZLF dataset elaboration. Patients under 18 and low-quality DFIs are excluded. Among the remaining DFIs, a total of 240 DFI of 232 unique patients were selected by active learning and manually segmented.}
	\label{fig:uzfsummary}
\end{figure}

Human data were obtained within the context of the study “Automatic glaucoma detection, a retrospective database analysis” (study number S60649). The Ethics Committee Research UZ/KU Leuven approved this study in November 2017 and waived the need for informed consent.
A total of 115,237 optic disc-centered DFIs from 13,185 unique patients, captured between 2010 and 2019 were provided by the UZL in Belgium. These DFIs were taken with a Visucam Pro NM camera with 30$^{\circ}$ FOV (Zeiss). The resolution of these DFIs was 1444x1444 pixels which is higher than most public DFIs datasets and enables the visualization of smaller blood vessels. The median age of the DFIs and the interquartiles (Q1-Q3) were 64 (52-75) years and $52\%$ were female. The excluded criteria included patients under 18 years of age and low-quality DFIs (FundusQ-Net $<6$) \cite{Abramovich2022FundusQ-Net:Grading} (Figure \ref{fig:uzfsummary}). Active learning was used to proactively select DFIs for manual segmentation. Specifically, among the extracted DFIs, an exploration-exploitation strategy was applied to select 240 DFIs for A/V segmentation. The exploration step consisted of randomly stratified sampling of some DFIs to annotate according to the patient's sex and the imaged eye (right / left). The exploitation steps, or active learning steps, consisted of selecting a set of DFIs with low LUNet A/V segmentation performance. Low segmentation performance was defined as a lack of vessel continuity. A subset of 240 DFIs denoted UZL Fundus (UZLF), from 232 unique patients, were selected and manually segmented. The patients included in UZLF were between 18 and 90 years of age (median (interquartile): 62 (50-73) years) and $58\%$ were female. UZLF was $57\%$ composed of left eye DFIs. Patients who belong to the UZLF dataset were separated into the following classes: (1) Normal ophthalmic findings (NOF), (2) Normal tension glaucoma (NTG), (3) Primary open angle glaucoma (POAG), and (4) other condition (OTHER). 

\subsubsection{Universidad Nacional de Asunción Fundus (UNAF) external test set}
The dataset of fundus images for the study of diabetic retinopathy contains 757 adult patient DFI acquired in the Department of Ophthalmology of the Hospital de Clínicas of San Lorenzo, Paraguay \cite{Benitez2021DatasetRetinopathy}. The DFIs were acquired using a  Visucam  500 camera with 45$^{\circ}$ FOV (Zeiss). The eyes included in this study were classified as: (1) No signs of diabetic retinopathy (NDR), (2) Mild or early, non-proliferative diabetic retinopathy (NPDR), (3) Moderate NPDR, (4) Severe NPDR, (5) Very severe NPDR, (6) Proliferative diabetic retinopathy (PDR), (7) Advanced PDR. The resolution of the original DFIs is 2124x2056 pixels. In order to benchmark LUNet on this dataset, the DFIs were padded by zeros to a squared resolution of 2124x2124 and then cropped to a 1444x1444 resolution. From the resulting DFIs, 15 optic disc-centered DFIs were randomly selected to form the UNAF external test set. The UNAF set included four NDR, two moderate NPDR, seven patients with severe NDPR, and two patients with very severe NDPR. No additional clinical data were available for this dataset.

\subsubsection{University of Iowa Hospitals and Clinics (INSPIRE-AVR) external test set}
This dataset contains 65 DFIs acquired from patients with POAG at the University of Iowa Hospitals and Clinics. DFIs were acquired using a 30$^{\circ}$ Zeiss fundus camera \cite{Niemeijer2011AutomatedPhotographs}. The images were centered on the optic disc. The original DFIs resolution was 2392x2048. In order to benchmark LUNet on this dataset, the black border of the DFIs were padded to a squared resolution of 2048x2048 pixels and then resized to a 1444x1444 pixels resolution. From the resulting DFIs, 15 optic disc-centered DFIs were randomly selected to form the second external test set. No other additional metadata were provided in the open source dataset. 

\subsubsection{LES-AV dataset}
This dataset contains 22 optic disc-centered DFIs with A/V segmentation acquired at the UZL in Belgium \cite{Orlando2018TowardsImages}. Of these, 21 were captured with a resolution of 1444x1620 and a field of view of 30$^{\circ}$. In order to benchmark LUNet on this dataset, the black borders of the 21 DFIs were cropped to a squared resolution of 1444x1444 pixels and was used as external test set. The DFI numbered 275 was excluded due to its inclusion in the UZLF dataset. This resulted in 20 images in LES-AV including DFIs from 10 healthy participants, 6 NTG patients and 4 POAG patients. Among the patients, $45\%$ were women and the median and interquartiles age of the DFIs were 71 (62-80). 

\subsubsection{HRF dataset}
This dataset contains 45 non optic disc-centered DFIs with A/V segmentation \cite{Budai2013RobustImages}. The original DFIs resolution was 3504x2336 with a field of view of 60$^{\circ}$. In order to benchmark LUNet on this dataset, we extracted 1444x1444
pixel ROI centered around the optic disc when possible.
This resulted in 28 DFI which were used as an additional
external test set. The eyes included in this study were classified as: (1) Healthy, (2) Glaucoma, and Diabetic retinopathy. No additional clinical data were available for this dataset.

\begin{table}[tb]
\begin{center}
\resizebox{\columnwidth}{!}{%
\begin{tabular}{|l|c| c| c| c| c |c |c| c|}
\hline
Name&N\textsuperscript{\underline{o}} DFIs & Country & Age & Original FOV & Purpose \\
\hline\hline
UZLF&240&Belgium&62 (50-73) & 30$^{\circ}$ &Train \& test\\
UNAF&15&Paraguay&-& 45$^{\circ}$& External test\\
INSPIRE-AVR&15&United State&-& 30$^{\circ}$& External test\\
LES-AV&20&Belgium&71 (62-80)& 30$^{\circ}$& External test\\
HRF (centered version)&28&Germany&-& 60$^{\circ}$& External test\\
\hline
\end{tabular}
}
\end{center}
\caption{Summary of the optic disc-centered DFI datasets with A/V segmentation. Median (Q1-Q3) are provided for Age.}
\label{table:1}
\end{table}

\subsection{Reference segmentations}
The INSPIRE-AVR, UZLF and UNAF datasets were manually segmented by the retinal experts of the UZ Leuven Hospital using the Lirot.ai app developed in our previous work \cite{Fhima2022Lirot.Segmentations} and following the protocol described in Fhima et al. \cite{Fhima2022PVBM:Segmentation}. 

A total of sixteen annotators experienced in micro-vascular research worked between July 2021 and January 2023 to build the UZLF dataset. The annotators were divided into two groups; (1) one experienced senior annotator, an ophthalmology resident with a PhD in retinal vascular biomarkers, and (2) 15 junior annotators who were all graduate students in medicine with a research internship of \textgreater1 month at the Research Group of Ophthalmology and trained by the senior annotator. The UNAF and INSPIRE-AVR datasets were annotated by the senior annotator. For the UZLF dataset, the junior annotators performed the first segmentation of the 240 DFIs, of which 174 were later corrected by the senior annotator to avoid mistakes and to correct for different annotation styles.

\subsection{Data preparation}
A DFI was modeled by a couple $(X,y)$, where:
\begin{itemize}
    \item $X \in R^{w \times h \times 3}$ represents a DFI with a width of $w$, a height of $h$ pixels and 3 channels, i.e. red, green and blue;
    \item $y \in R^{w \times h \times 3}$ where y = [$y_{a}$, $y_{v}$, $y_{ukn}$] with $y_{a}$,$y_{v}$,$y_{ukn}$ $\in R^{w \times h \times 1}$, i.e., $y_{a}$, $y_{v}$ and $y_{ukn}$ are binary images., where $y_{a}$ represent the indistinguishable blood vessels, with white pixel if part of a blood vessel, and black pixel otherwise. 
\end{itemize}
Where $h=w=1444$.

\subsubsection{Train-Validation-Test set elaboration}
The test set was elaborated by randomly selecting 14 patients from each of the POAG, NTG and Normal subgroups from the 174 expert-reviewed segmentations and by selecting $50\%$ women for each subgroup. This resulted in a test set of 50 DFIs from 42 patients. In addition, a total of 6 DFI randomly selected among ophthalmology diseases were manually segmented and added to the test set to evaluate the generalization performance of LUNet for other diseases. Among the remaining DFIs, i.e. 184 images, $85 \%$ were used for training and $15 \%$ for validation while images were stratified by patient thus ensuring no information leakage.

\subsubsection{Preprocessing}
LUNet is based on convolutions with same padding and max-pooling 2D (2x2). LUNet uses six levels of depth, which means that the height and width of the input are divided by 2 six times. Thus, LUNet's input size needs to be divisible by $2^6$. To address this requirement, zero-padding is applied to the inputs and outputs to obtain an input tensor of size 1472x1472x3 and an output tensor of size 1472x1472x2. Furthermore, the input and output images were normalized.

\subsection{DL Architecture}
\begin{figure*}[!tb]
    \centering
	\includegraphics[width=\textwidth]{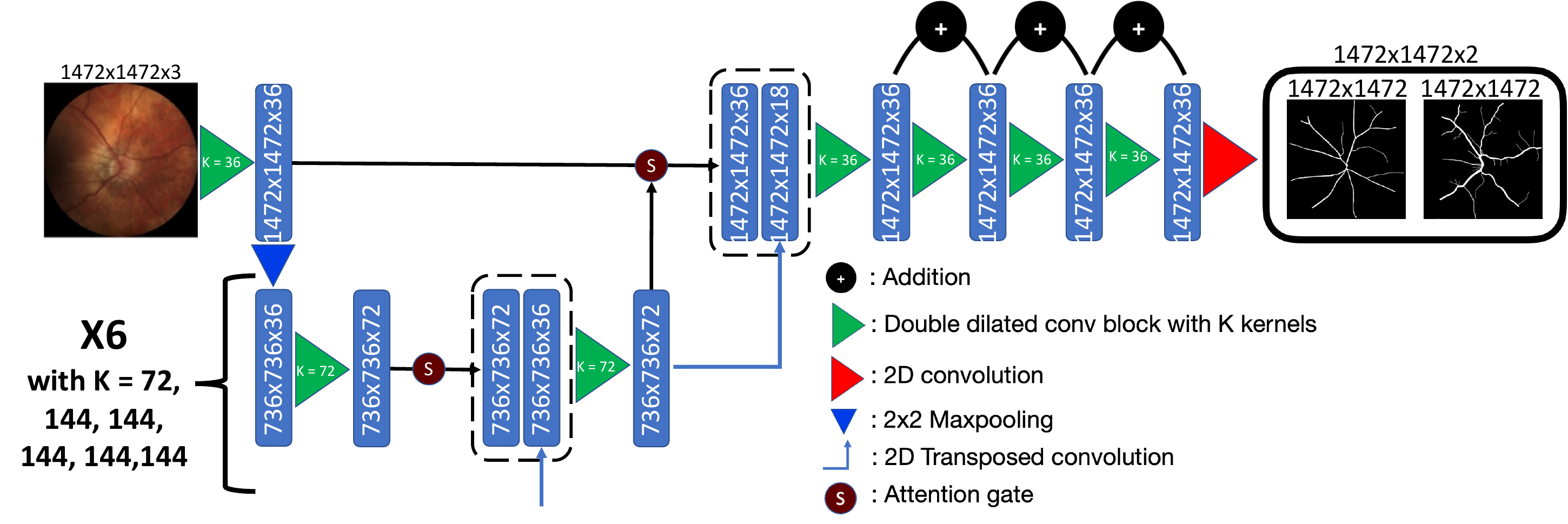}
	\caption{LUNet architecture based on an attention U-Net backbone with double dilated convolution blocks, an increase depth, a long tail and an over representation of the features extracted from the encoder compared to the one extracted from the decoder at each level of depth.}
	\label{fig:archi}
\end{figure*}

The attention U-Net architecture was used as the backbone for our DL algorithm \cite{Oktay2018AttentionPancreas}. 
The A/V segmentation task is challenging due to the long-range dependencies of the blood vessels and their small width. To achieve accurate A/V segmentation, it is necessary to have a model that can identify both local and global dependencies. The local dependencies help detect the smaller blood vessels more accurately, while the global dependencies enable the reconstruction of the full blood vessel. In convolution neural networks (CNN), the local dependencies are captured by the convolution operation, while the larger range dependencies are dependent on the receptive field of the CNN. The receptive field can be increased by augmenting the depth of the CNN and increasing the kernel size of the convolutional layers, using dilated convolution, or using some max pooling layers. The first two solutions would lead to an augmentation of the model parameters with a small increase of the receptive field, while the max pooling would not add any parameter and would lead to a much larger receptive field. Nevertheless, due to the small size of the blood vessel in a classical DFI, the max pooling could be an excessively aggressive strategy, which would result in the loss of small-blood-vessel visibility after several iterations. To tackle this problem, LUNet was designed to include several improvements, including; (1) a new double dilated convolution block with an increased receptive field, (2) a long tail that works on the full pixel resolution, (3) an increased depth, (4) an over-representation of the feature extracted by the encoder compared to the one reconstructed by the decoder at each level of depth of the LUNet auto-encoder. The LUNet architecture is shown in Figure \ref{fig:archi}.

\subsubsection{Double dilated convolution block}
\begin{figure}[!tb]
    \centering
	\includegraphics[width=0.35\textwidth]{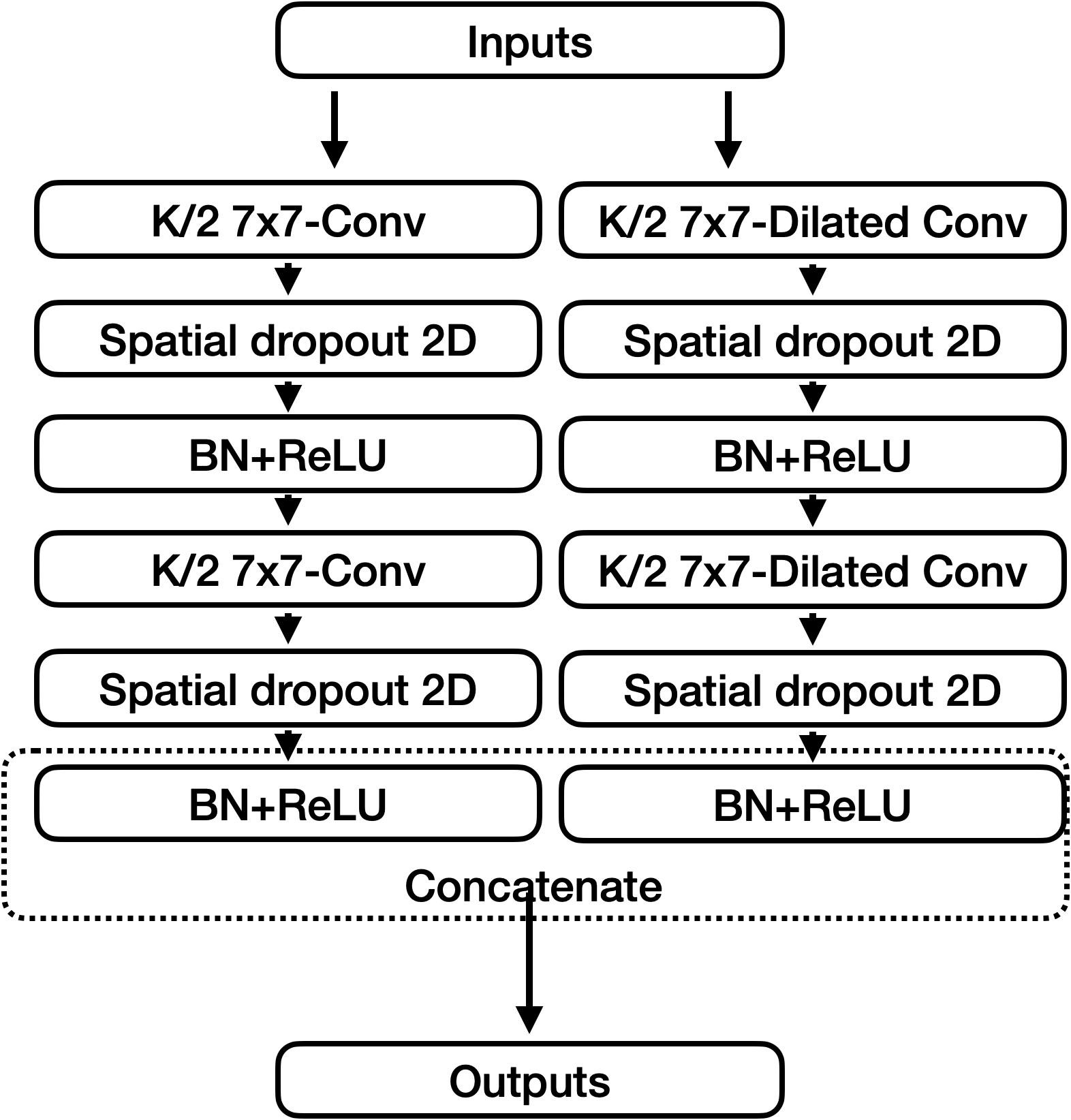}
	\caption{LUNet double dilated convolution block with parameter K.}
	\label{fig:ddcb}
\end{figure}

Dilated convolution enables a larger receptive field but would lead to a less accurate feature representation of small details (due to the dilation rate). To tackle this limitation, both classical and dilated convolutions with a kernel size of 7 were used in the model. The double dilated convolution block also incorporated a spatial dropout 2D regularization \cite{Tompson2015EfficientNetworks} and a batch normalization layer with the ReLU activation function. A diagram of the double dilated convolution block is presented in Figure \ref{fig:ddcb}.

\subsubsection{Long tail}
Classical convolutional neural network architecture uses max pooling to increase receptive fields and obtain a larger long-range dependency. Nevertheless, some blood vessels may be too small to remain detectable after one or more max poolings. To tackle this limitation, four double-dilated convolution blocks were added at the end of LUNet to: (1) perform several computations at the full image resolution, to increase the possibility of detection of small blood vessels that may not be detectable at lower resolutions; (2) Increase the receptive field of the full-resolution-computed features. Overall the long tail helps to refine the segmentation.

\subsection{Loss}
Other approaches have described the A/V segmentation problem as a multiclass segmentation problem leading to a multiclass cross-entropy loss minimization \cite{Hu2021AutomaticImages,Hemelings2019ArteryveinNetwork}.
We formulated the problem differently, as a binary multi-label segmentation because it is common to see superimposition of arterioles and venules on several pixels of a DFI. Furthermore, some manual segmentations may contain pixels labeled as unknown. These pixels were not penalized for being classified as A or V. However, LUNet was still trained to detect these pixels and eventually classify them. A regularization term was added to the loss function to learn to detect blood vessels without A/V distinction. Accordingly, the LUNet loss function, $L_{LUNet}$, was defined as the sum of three losses computed independently on the arterioles, the venules, and the overall blood vessels.
\begin{equation}
\begin{aligned}
L_{LUNet}(y,\hat y) = L(y_{a} \times (1 -y_{ukn} ),\hat y_{a}\times (1 -y_{ukn} )) \\+ L(y_{v}\times (1 -y_{ukn} ),\hat y_{v}\times (1 -y_{ukn} )) \\
+ L(max(\hat y_{a},\hat y_{v}),max(y_{a},y_{v},y_{ukn}))
\end{aligned}
\end{equation}
with
\begin{equation}
    \begin{aligned}
 L(y,\hat y) = \lambda_1 \times L_{BCE}(y,\hat y) + \lambda_2 \times L_{dice}(y,\hat y) + \\ \lambda_3 \times L_{clDice}(y,\hat y) +  \nabla (\hat y)
    \end{aligned}
\end{equation}

where $y$ is the ground truth segmentation, $\hat y$ is the predicted probability map, $L_{BCE}$ is the binary cross-entropy loss, $L_{dice}$ is the dice loss, $L_{clDice}$ is the centerline dice loss \cite{Shit2020ClDice-aSegmentation.} and where the gradient of $\hat y$ is minimized to favor the continuity of the blood vessels by constraining the change of value in the probability map to be smoother. 

\subsection{Training protocol}
LUNet hyperparameters were manually fine-tuned using the validation set. LUNet was trained for up to 1300 epochs, while retaining the model which obtains the minimum validation loss. A batch size of 8 and an Adam optimizer with $1e^{-4}$ for the learning rate were used. The $L_{LUNet}$ loss function was used with $\lambda_1 = \lambda_2 = 1$ and $\lambda_3 = 0.3$.

For each training DFI loaded, random online data augmentation was performed with (1) horizontal and vertical flip, (2) transposition, (3) rescaling of the input and output to a lower resolution uniformly sampled between 800x800 and 1472x1472, and (4) color jittering.
Test time data augmentation was used for the test set prediction \cite{Wang2018Test-timeSegmentation} with the following rotation angles [0, 30, 60, 90, 120, 150, 180, 210, 240, 270, 300, 330], as well as transposition of the DFI which resulted in 24 predicted segmentations for a single original DFI. The inverse transform for each prediction was performed. The final predicted probability map for the segmentation was the pixel-wise average of all the 24 segmentations.

\subsection{Benchmark}

\textit{Benchmark against SOTA algorithms:} LUNet performance on the UZLF dataset was compared to that of BFN \cite{Zhou2021LearningImaging} and VC-Net \cite{Hu2021AutomaticImages}, two SOTA algorithms, that were trained on the UZLF-train dataset. On the external datasets, LUNet was compared to VC-Net \cite{Hu2021AutomaticImages} and BFN \cite{Zhou2021LearningImaging} trained on the UZLF dataset as well as the publicly trained version of BFN* \cite{Zhou2022AutoMorph:Pipeline}.

\textit{Benchmark against junior performance:} The DFIs of the UZLF-test dataset were annotated by a junior annotator and then reviewed by the senior annotator. This enables a comparison of the dice score between an individual junior annotator and the senior annotator. This dice score statistic allows to benchmark LUNet to the performance of a human junior annotator. 

\subsection{Performance measures}
A dice score was computed for the arterioles segmentation ($dice_a$) and for the venules segmentation ($dice_v$). In order to estimate the 95\% confidence interval (CI), a bootstrap method was employed by repeatedly sampling 80\% of the test set with replacement, and computing the mean score of each sample. This procedure was carried out 1000 times, and the resulting distribution of means was used to determine the lower and upper bounds of the 95\% confidence interval. In order to monitor the performance of LUNet with respect to the number of annotated DFIs, two learning curves one for the $dice_a$ and one for the $dice_v$ were computed. The reported performance was computed on the UZLF-test set as the training set size increased. For the LES-AV and HRF datasets, some of the ground truth blood vessels were annotated as unknown. The unknown pixels were not taken into account in the computation of the two dice scores.

Finally, LUNet was evaluated on the local test set with respect to the ability to correctly estimate vasculature biomarkers (VBMs). VBMs were computed using the open source PVBM toolbox (\href{https://pvbm.readthedocs.io/}{https://pvbm.readthedocs.io/}) \cite{Fhima2022PVBM:Segmentation}. The VBMs include: the area (AREA), the tortuosity index (TI), the median arc-chord tortuosity (TOR), the length (LEN), the median branching angles (BA), the number of blood vessels that intersect with the optic disc (START), the number of endpoints (END), the number of intersection points (INTER), the fractals dimension (D0, D1 and D2). Each biomarker was computed independently on the arterioles and the venules and the Pearson correlation between the ground truth and the estimated VBMs were computed. The average Pearson correlation between all the VBMs was also computed in order to provide an overall performance measure. 


\section{Results}
\subsection{Performance on the UZLF-test}
\begin{figure}[tb]
    \centering
	\includegraphics[width=0.5\textwidth]{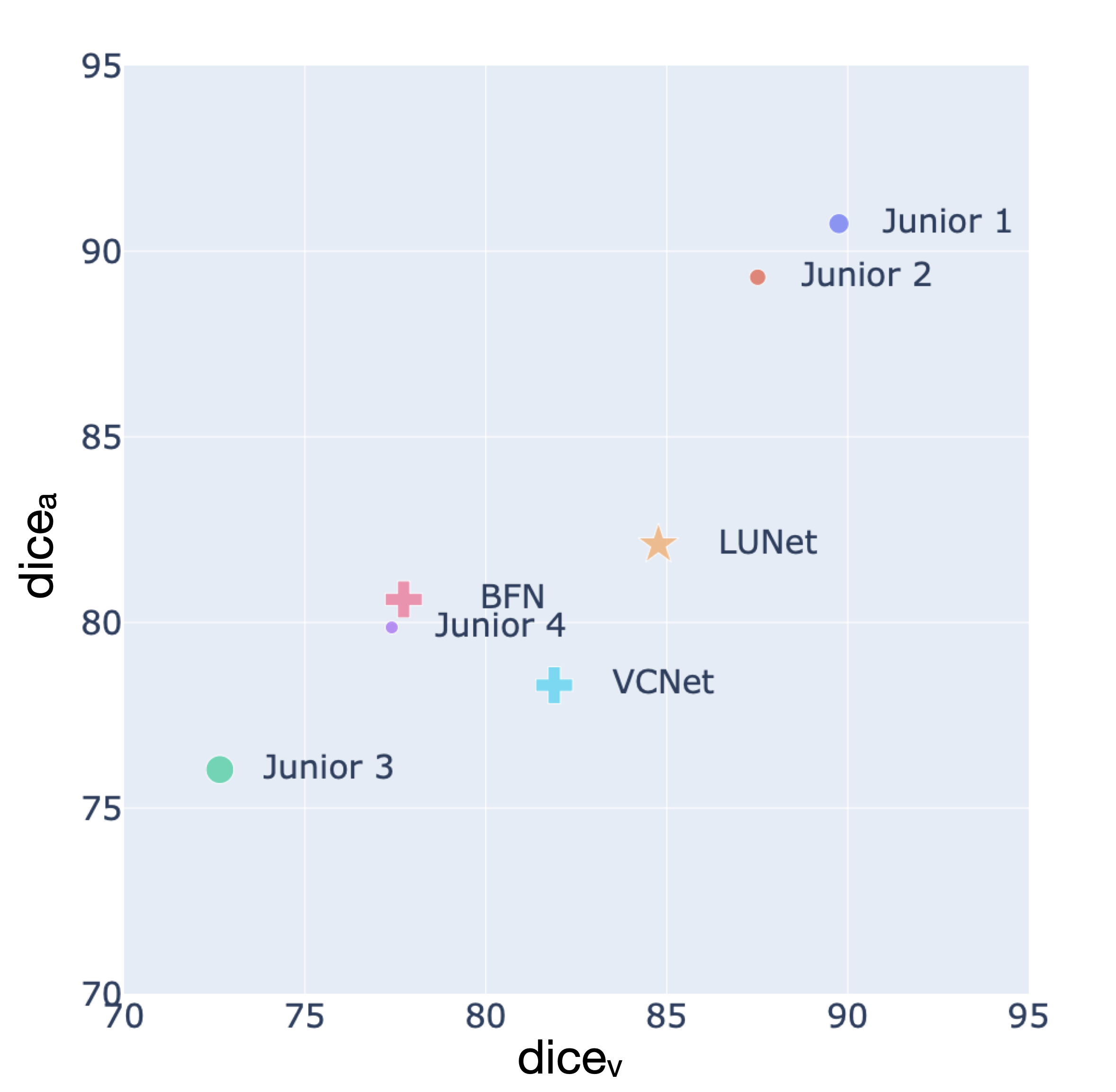}
	\caption{Performance of models and humans for A/V segmentations ($dice_a$ and $dice_v$) on the UZLF-test. The results for LUNet and benchmark SOTAs BFN \cite{Zhou2021LearningImaging} and VC-Net \cite{Hu2021AutomaticImages} are shown. Performance results of junior annotators who segmented at least 5 DFIs on the UZLF-test are also shown. For the junior annotators, the size of the marker is proportional to the number of DFIs that they annotated in UZLF-test. LUNet performs better than the other DL algorithm and has comparable performance to an average junior annotator.}
	\label{fig:LUNetvsHuman}
\end{figure}

\begin{figure}[tb]
    \centering
    \includegraphics[width=0.5\textwidth]{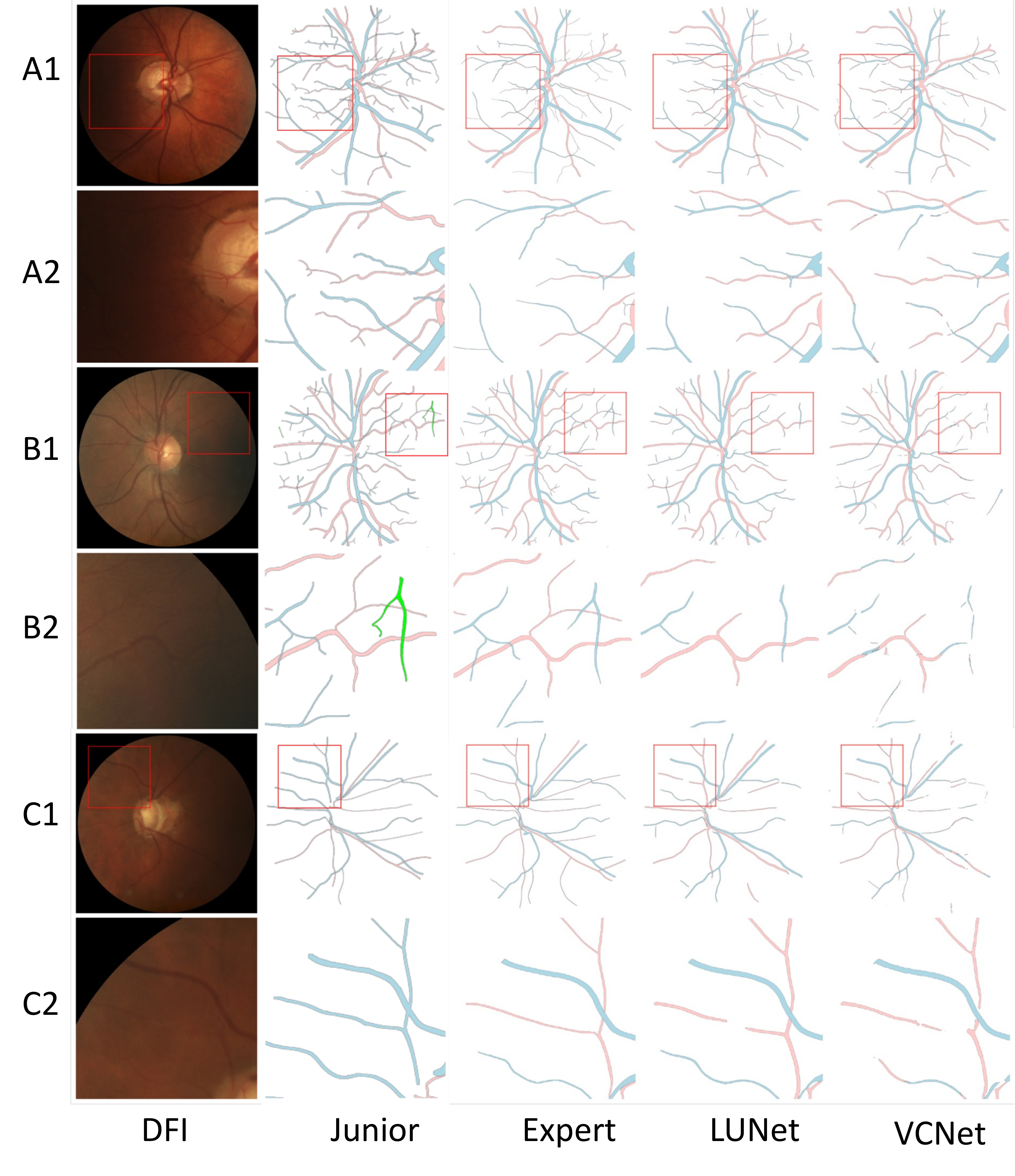}
    \caption{Examples of segmentations on the UZLF dataset. Segmentations performed by the 
    junior annotator, expert annotator, LUNet and VC-Net \cite{Hu2021AutomaticImages}. A1/A2: DFI with the best dice scores for LUNet, B1/B2: DFI with an intermediate LUNet dice scores, C1/C2: DFI with the worst LUNet dice scores.}
    \label{fig:LUNetResults}
\end{figure}

\begin{figure}[tb]
    \centering
	\includegraphics[width=0.4\textwidth]{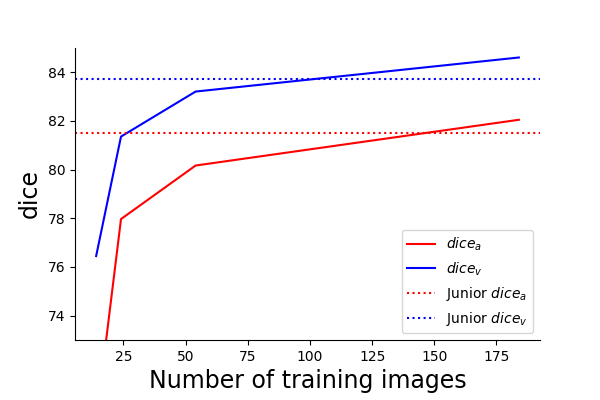}
	\caption{(A): LUNet's learning curves. Dice scores are shown for the UZLF-test set. The junior dice scores correspond to the average dice score of the junior annotators for the test set.}
	\label{fig:learningcurves}
\end{figure}

LUNet achieved dice scores of 81.99/84.54 for A/V segmentations on the local test set. These were higher than those of the SOTA benchmark algorithms VC-Net \cite{Hu2021AutomaticImages} and BFN \cite{Zhou2021LearningImaging} which achieved dice scores of 78.31/81.89 and 78.14/80.79, respectively. The performance of LUNet were better for A (81.99 vs. 81.50) and for V (84.54 vs 83.74) than an average junior annotator. Table \ref{table:localtest} and Figure \ref{fig:LUNetvsHuman} summarizes the quantitative results on the UZLF-test. The learning curves of LUNet are shown in Figure \ref{fig:learningcurves}. Figure \ref{fig:LUNetResults} shows examples of segmentations performed by LUNet, a junior annotator, an expert annotator, and the highest performing SOTA VC-Net \cite{Hu2021AutomaticImages}. Specifically, the DFI with the best, intermediate and worst LUNet performance was selected. Furthermore, LUNet outperforms VC-Net and BFN in estimating most VBMs (Table \ref{tab:VBMS}). LUNet obtained the best Pearson correlation score for 9 over 22 VBMs, while the junior annotators obtained the best performance for 9 other VBMs.

\begin{table}[tb]
\begin{center}
\resizebox{\columnwidth}{!}{%
\begin{tabular}{|l|c|c|}
\hline
Model&\multicolumn{2}{c|} {UZLF-test}\\
& \multicolumn{1}{c}{$dice_a$} & $dice_v$ \\ 

\hline\hline
BFN \cite{Zhou2021LearningImaging} &78.14 (76.31-79.66)
&80.79 (79.49-81.90)
\\
VC-Net \cite{Hu2021AutomaticImages} &78.31 (76.79-79.74)
&81.89 (80.95-82.93)
\\
Junior& 81.50 (78.00-84.41)
&83.74 (80.91-86.28)
\\
LUNet (our work) &\textbf{81.99 (80.63-83.11)
}&\textbf{84.54 (83.59-85.42)
}\\
\hline
\end{tabular}
}
\end{center}
\caption{Quantitative A/V segmentation results on the local (UZLF-test) test sets, including the average performance and (CI) 
for the $dice_a$ and $dice_v$. The table includes the performance of the junior annotators, LUNet, BFN \cite{Zhou2021LearningImaging} and VC-Net \cite{Hu2021AutomaticImages} trained on the UZLF-train dataset and evaluated on the UZLF-test dataset.}
\label{table:localtest}

\end{table}

\begin{table*}[tb]
\begin{center}
\resizebox{\textwidth}{!}{%
\begin{tabular}{|l|c|c|c| c| c| c|c|c|}
\hline
Model&\multicolumn{2}{c|} {LES-AV}&\multicolumn{2}{c|} {UNAF}&\multicolumn{2}{c|} {INSPIRE-AVR}&\multicolumn{2}{c|} {Cropped HRF}\\
& \multicolumn{1}{c}{$dice_a$} & $dice_v$& \multicolumn{1}{c}{$dice_a$} & $dice_v$& \multicolumn{1}{c}{$dice_a$} & $dice_v$&\multicolumn{1}{c}{$dice_a$} & $dice_v$ \\ 
\hline\hline
\multirow{2}{*}{
BFN* \cite{Zhou2021LearningImaging} }
&\multirow{2}{*}{
--}
&\multirow{2}{*}{
--}
&64.13 &72.78 &62.98 &68.10&70.95& 75.47 \\

&&&(59.93-68.07)&(70.24-74.98)&(57.41-66.88)&(63.06-71.30)&(68.73 - 73.26)&(73.75 - 77.12)\\
\multirow{2}{*}{
BFN \cite{Zhou2021LearningImaging}}
& 78.68 & 80.34 &68.01 &72.67 &68.61 &71.92& 72.90 &76.99\\
&(76.34-80.98)&(78.85-81.75)&(63.45-71.92)&(69.67-75.47)&(66.23-70.84)&(67.74-75.04)&(70.52-75.37)&(75.74-78.38)\\
\multirow{2}{*}{
VC-Net \cite{Hu2021AutomaticImages}}
& 78.75 & 82.49 &68.71 &74.43 &71.28 &75.34&75.78& 79.99 \\
&(76.19-81.11)&(80.82-84.17)&(64.91-71.86)&(72.72-76.19)&(69.50-72.93)&(74.13-76.60)& (73.77-77.55)& (79.03- 81.04)\\

\multirow{2}{*}{
LUNet (Our work) }
&\textbf{82.30}&\textbf{84.75}&\textbf{73.31 }&\textbf{79.04 }&\textbf{73.58 }&\textbf{77.53 }&\textbf{78.12} &\textbf{80.39}\\
&\textbf{(80.26-84.43)}&\textbf{(83.02-86.28)}&\textbf{(70.69-76.41)}&\textbf{(77.84-80.39)} & \textbf{(70.94-75.69)}&\textbf{(74.99-79.84)}&\textbf{(76.10- 80.11)}&\textbf{(79.20-81.72)} \\
\hline
\end{tabular}
}
\end{center}
\caption{Performance in A/V segmentation for the external test sets, including the average performance and (CI) for the $dice_a$ and $dice_v$. The table includes LUNet, BFN \cite{Zhou2021LearningImaging} and VC-Net \cite{Hu2021AutomaticImages} trained on the UZLF-train dataset. It also include the performance of BFN* \cite{Zhou2022AutoMorph:Pipeline} publicly available version, i.e. pretrained by the original authors work. Results for BFN* are not provided for LES-AV because the model was originally trained on this dataset.}
\label{table:extset}
\end{table*}

\begin{table}[tb]
\begin{center}
\resizebox{\columnwidth}{!}{%
\begin{tabular}{|l|c|c|c|c|c|}
\hline
 & LUNet & VC-Net & BFN&Junior \\ \hline
AREA\_a & 0.823 & \textbf{0.837} & 0.833&0.744  \\
TI\_a & 0.922 & 0.888 & 0.932&\textbf{0.935}  \\
TOR\_a & 0.796 & 0.713 & \textbf{0.799}&0.752 \\
LEN\_a & \textbf{0.811} & 0.770 & 0.757 & 0.796  \\
BA\_a & 0.711 & 0.512 & 0.640&\textbf{0.798 } \\
START\_a & \textbf{0.685} & 0.561 & 0.581&0.677  \\
END\_a & 0.806 & 0.690 & 0.731&\textbf{0.869}  \\
INTER\_a & 0.811 &0.644 & 0.732 & \textbf{0.887 } \\
D0\_a & \textbf{0.867} & 0.850 & 0.863& 0.805  \\
D1\_a & \textbf{0.875} & 0.867 & 0.856&0.810  \\
D2\_a & \textbf{0.867} & 0.865 & 0.856& 0.808 \\
AREA\_v & \textbf{0.883} & 0.878 & 0.869 & 0.773  \\
TI\_v & 0.884 & 0.812 & \textbf{0.888} & 0.847 \\
TOR\_v & \textbf{0.861} & 0.840 & 0.803& 0.761  \\
LEN\_v & 0.771 & 0.557 & 0.758& \textbf{0.822}  \\
BA\_v & 0.370 & 0.245 & 0.442 & \textbf{0.480} \\
START\_v & 0.671 & 0.486 & 0.442& \textbf{0.741}  \\
END\_v & 0.785 & 0.728 & 0.731 & \textbf{0.882} \\
INTER\_v & 0.731 & 0.627 & 0.662& \textbf{0.838} \\
D0\_v & 0.864 & 0.822 & \textbf{0.871} & 0.783 \\
D1\_v & \textbf{0.862} & 0.816& 0.813&0.730 \\
D2\_v & \textbf{0.848} & 0.805 & 0.794&0.700 \\
\hline

\end{tabular}
}
\caption{Pearson correlation  between the ground truth and estimated VBM based on a given segmentation algorithm. VBMs are computed independently for arterioles (a) and for venules (v).}
\label{tab:VBMS}
\end{center}
\end{table}

\subsection{Generalization performance}
\begin{figure}[tb]
    \centering
    \includegraphics[width=0.5\textwidth]{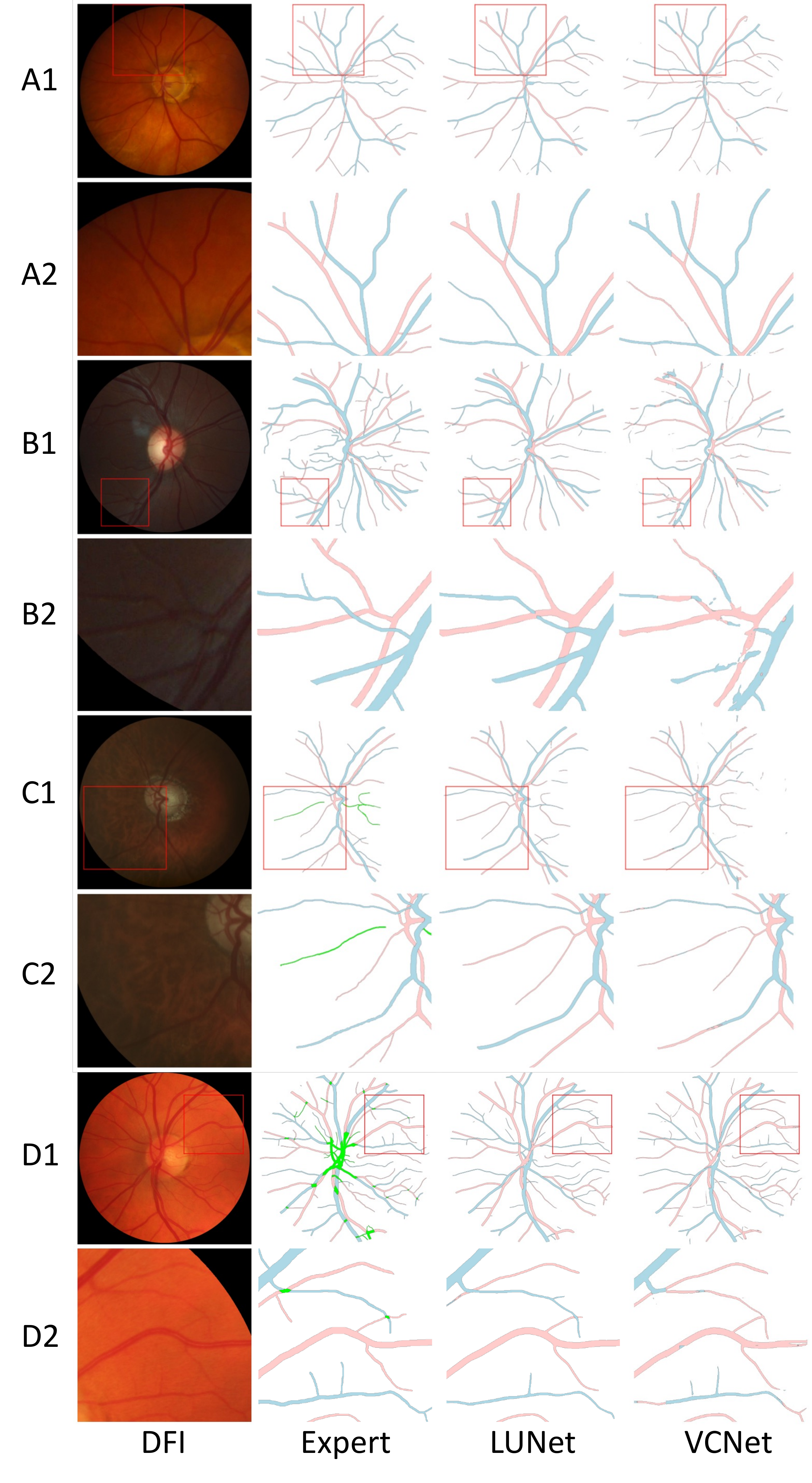}
    \caption{Examples of segmentations on the external datasets. Segmentations performed by the expert annotator, LUNet and VC-Net \cite{Hu2021AutomaticImages} trained on the UZLF dataset. A1/A2: DFI in the INSPIRE-AVR dataset, B1/B2: DFI in the UNAF dataset, C1/C2: DFI in the LES-AV dataset and D1/D2: DFI in the cropped HRF dataset.}
    \label{fig:extquali}
\end{figure}

Figure \ref{fig:extquali} shows the segmentations generated by LUNet, VC-Net \cite{Hu2021AutomaticImages}, and BFN \cite{Zhou2021LearningImaging} performance for a DFI example for each of the external test sets. Table \ref{table:extset} summarizes the quantitative results on the external test sets.


LUNet achieved average dice scores of 82.30/84.75 for A/V segmentation on the LES-AV dataset. This performance was superior to those of the SOTA benchmark algorithms VC-Net \cite{Hu2021AutomaticImages} and BFN \cite{Zhou2021LearningImaging} which achieved dice scores of 78.75/82.49 and 78.68/80.34, respectively. The public version of BFN* \cite{Zhou2022AutoMorph:Pipeline} was not evaluated on LEV-AV because it was originally trained on it. LUNet achieved dice scores of 73.31/79.04 for A/V segmentations on the UNAF dataset. This performance was superior to those of the SOTA benchmark algorithms VC-Net \cite{Hu2021AutomaticImages} and BFN \cite{Zhou2021LearningImaging} which achieved dice scores of 68.71/74.43 and 68.01/72.67, respectively. Its performance was also higher than that of the public version of BFN* \cite{Zhou2022AutoMorph:Pipeline} which achieved A/V dice scores of 64.13/72.78. LUNet achieved dice scores of 73.58/77.53 for A/V segmentations on the INSPIRE-AVR dataset. These were higher than the SOTA benchmark algorithms VC-Net \cite{Hu2021AutomaticImages} and BFN \cite{Zhou2021LearningImaging} which achieved dice scores of 71.28/75.34 and 68.61/71.92, respectively. They were also higher than those of the public version of BFN* \cite{Zhou2022AutoMorph:Pipeline} which achieved A/V dice scores of 62.98/68.10. LUNet achieved dice scores of 78.12/80.39 for A/V segmentations on the cropped HRF dataset. This performance was superior to those of the SOTA benchmark algorithms VC-Net \cite{Hu2021AutomaticImages} and BFN \cite{Zhou2021LearningImaging} which achieved dice scores of 72.90/76.99 and 75.78/79.99, respectively. Its performance was also higher than that of the public version of BFN* \cite{Zhou2022AutoMorph:Pipeline} which achieved A/V dice scores of 70.95/75.47. 

\subsection{Ablation study}
\begin{figure}[tb]
    \centering
    \includegraphics[width=0.5\textwidth]{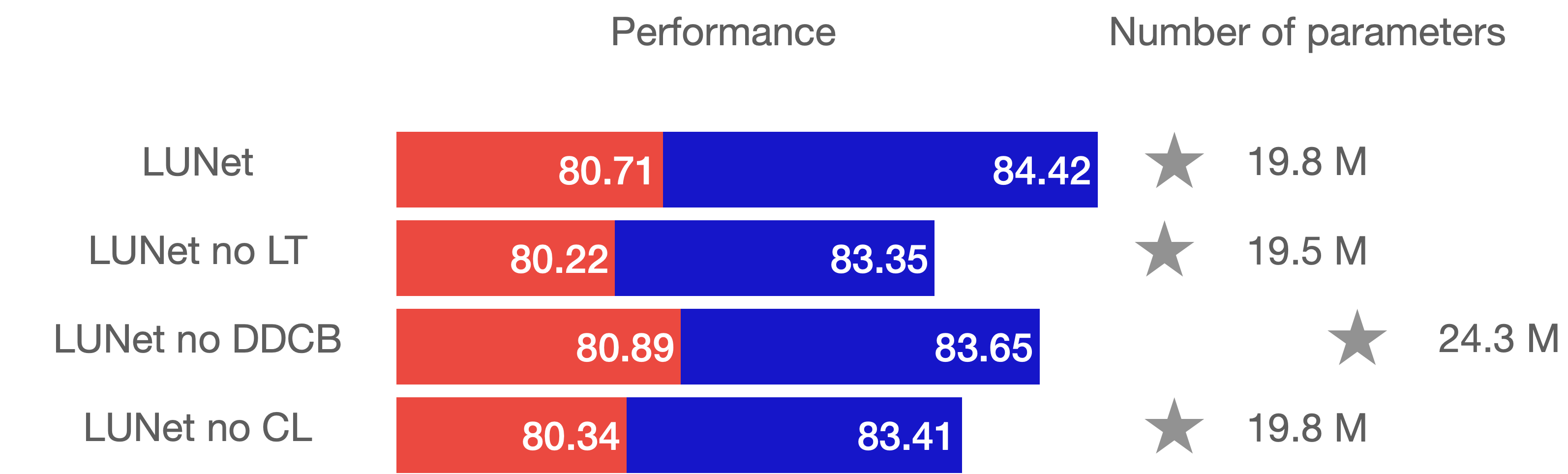}
    \caption{Ablation study. Results are reported for the validation set.}
    \label{fig:ab1}
\end{figure}

An ablation study was conducted on the validation set by imputing the following architecture components, LT: long tail, CL: custom loss, DDCB: double dilated convolution block (Figure \ref{fig:ab1}). This ablation study demonstrates the importance of the individual components.

\section{Discussion and future work}
The first main contribution of this work is the creation of UZLF, a new dataset of high resolution DFIs with A/V segmentation which is six times larger than existing open datasets. Lirot.ai, with an active learning pipeline, was used to optimize the quality-time trade-off in developing the UZLF dataset. Additionally, 30 DFIs from the UNAF and INSPIRE-AVR public datasets were A/V segmented. The new DFI dataset with reference A/V segmentations is made open access (upon publication).

Our second main contribution is the development of a novel and robust DL model, denoted as LUNet, for the automated segmentation of A/V based on optic disc-centered and high-resolution DFIs. LUNet outperformed two open source SOTA algorithms, namely  VC-Net \cite{Hu2021AutomaticImages} and BFN \cite{Zhou2021LearningImaging,Zhou2022AutoMorph:Pipeline}. LUNet performance (81.99/84.54 for A/V) was comparable to an average trained junior annotator (81.50/83.74 for A/V) on the local test set. There was no significant difference in performance on the local test set for male vs female DFIs (A/V p-values are 0.83/0.34), left vs right eye DFIs (A/V p-values are 0.55/0.62) and POAG vs non POAG DFIs (A/V p-values are 0.68/0.59). LUNet consistently generalized better than the benchmarked algorithms for all four external test sets. However, performance dropped on the external test sets versus the local test set. Several factors can affect the rendering of a DFI and thus the performance of LUNet on the external test set. One factor that can contribute to the drop in performance is the lower quality of the DFI. FundusQ-Net \cite{Abramovich2022FundusQ-Net:Grading} referred a lower quality for the DFIs issued from the UNAF dataset with a median (Q1-Q3) of 7.24 (6.74 - 7.60) while the quality observed for the UZLF test set was 7.86 (7.51-8.38). Additionally, the original framing of the images in the UNAF dataset differs from the one in the UZLF dataset, which can affect the model performance. The quality of the INSPIRE-AVR dataset was slightly higher than the one of the UZLF with a median (Q1-Q3) FundusQ-Net score of 8.30 (7.77-8.45) and thus cannot explain the drop in performance observed for this external test set. Another possible reason is the difference in the population studied in the UNAF and INSPIRE-AVR datasets compared to the UZLF dataset. The UNAF datasets have a more significant number of patients with diabetic retinopathy, which can lead to distribution shifts related to the presence of hemorrhage and/or exudates. Furthermore, UNAF and INSPIRE-AVR datasets come from different countries; Paraguay for UNAF and United States for INSPIRE-AVR. The ethnic differences may also affect the performance of the model. Different ethnicities can lead to different variations in the retinal vasculature and overall retinal structure and pigmentation to which the model has to adapt \cite{Li2013RacialAsians}. The acquisition protocol used to capture the images can also play an important role in the performance drop. Factors such as the amount of light in the room, the quality of the equipment used, and the skill of the operator capturing the image can all affect the appearance of the images and, thus, the model’s generalization performance.  Overall, the drop in performance on the UNAF dataset may come from a different original framing, a lower quality of the DFIs and a different population sample in terms of pathology and ethnicity.
There are some limitations to this research and opportunities for improvement. Although LUNet performed better than other DL models on external datasets, there was a significant drop in performance compared to the local test set. This drop in performance could be reduced by training LUNet on a more diverse dataset using either supervised or self-supervised learning. Despite the relatively high number of manually segmented DFI, performance can continuously improve with the training set size, as shown in Figure \ref{fig:learningcurves}. This suggests that increasing the training set size further may lead to improved performance for LUNet. 

\section*{Conclusion} In conclusion, we developed LUNet a robust, i.e. high performing and generalizable, deep learning model for the segmentation of venules and arterioles in fundus images. We demonstrate how LUNet can be used to estimate vasculature biomarkers thus enabling large scale research on the effect of cardiovascular diseases on the eye vasculature.

\clearpage
\bibliographystyle{IEEEtran}
\bibliography{ieee_main}

\end{document}